# 3D MHD Simulations of Laboratory Plasma Jets

A. Ciardi[1,2], S.V. Lebedev[2], A. Frank[3,4], E. G. Blackman[3,4], D.J. Ampleford[6], C.A. Jennings[6], J.P. Chittenden[2], T. Lery[5], S.N. Bland[2], S.C. Bott[2], G.N. Hall[2], J. Rapley[2], F. A. Suzuki Vidal[2], and A. Marocchino[2].

[1]LUTH, Observatoire de Paris et UMR 8102 du CNRS, 92195 Meudon, France.
[2]The Blackett Laboratory, Imperial College, London SW7 2BW, UK
[3]Department of Physics and Astronomy, University of Rochester, Rochester NY USA
[4]Laboratory for Laser Energetics, University of Rochester, Rochester NY USA
[5]Dublin Institute for Advanced Studies, Dublin, Ireland
[6]Sandia National Laboratory, Albuquerque, New Mexico, USA

## ABSTRACT

Jets and outflows are thought to be an integral part of accretion phenomena and are associated with a large variety of objects. In these systems, the interaction of magnetic fields with an accretion disk and/or a magnetized central object is thought to be responsible for the acceleration and collimation of plasma into jets and wider angle flows. In this paper we present three-dimensional MHD simulations of magnetically driven, radiatively cooled laboratory jets that are produced on the MAGPIE experimental facility. The general outflow structure comprises an expanding magnetic cavity which is collimated by the pressure of an extended plasma background medium, and a magnetically confined jet which develops within the magnetic cavity. Although this structure is intrinsically transient and instabilities in the jet and disruption of the magnetic cavity ultimately lead to its break-up, a well collimated, "knotty" jet still emerges from the system; such clumpy morphology is reminiscent of that observed in many astrophysical jets. The possible introduction in the experiments of angular momentum and axial magnetic field will also be discussed.

# INTRODUCTION

The formation and collimation of jets is a problem of great interest in astrophysics. Jets are observed in a diversity of often unrelated systems and range from the sub-parsec and parsec scale in the case of young stellar object jets (Reipurth and Bally, 2001) to the galactic scale jets, thought to be powered by super-massive black holes present in the centre of active galactic nuclei (see Begelman *et al.*, 1984). Jets may also play a critical role in the formation of gamma-ray bursts (see Piran, 2005 for a review) and by association supernovae (Galama *et al.*, 1998; Stanek *et al.*, 2003; LeBlanc and Wilson, 1970; Khokhlov *et al.*, 1999; MacFadyen and Woosley, 1999; Wheeler *et al.*, 2002; Akiyama *et al.*, 2003). Such diversity, in otherwise similar outflow structures may suggest the presence of a universal formation mechanism and over the last twenty years, magnetic fields and the occurrence of rotation have been identified as the principal agents for creating collimated outflows. A common feature of the many variations of the magneto-rotational scenario (Blandford and Payne, 1982; Pudritz and Norman, 1986; Pelletier and Pudritz, 1992; Wardle and Koenigl, 1993; Shu *et al.*, 1994; Ustyugova *et al.*, 1999; Ustyugova *et al.*, 2000; Uchida and Shibata, 1985; Contopoulos and Lovelace, 1994; Ouyed *et al.*, 1997; Goodson *et al.*, 1999; Goodson and Winglee, 1999; Kudoh *et al.*, 2002), is that a magnetic field can extract the rotational energy and launch the plasma from a gravitational potential well to escape velocities. In many of the models, the winding of an initially poloidal magnetic field results in a flow pattern dominated by a toroidal magnetic field. In this context the interaction of a dominant toroidal magnetic field with thermal ambient plasma has been investigated in the laboratory both experimentally and numerically (Lebedev *et al.*, 2005b, Ciardi *et al.*, 2005). One of the aims of the present work is to extend the numerical work to 3D MHD simulations and to investigate the late stages of the evolution of "magnetic tower" in the laboratory.

Laboratory experiments, performed on a variety of high energy density facilities, are starting to address some important astrophysical issue (Remington, 2005).

Conical wire arrays have been successfully used to produce radiatively cooled, hypersonic jets (Ciardi *et al.*, 2002; Lebedev *et al.*, 2002) and to study their interaction with an ambient medium (Lebedev *et al.*, 2004; Ampleford *et al.*, 2005; Lebedev *et al.*, 2005a). In these experiments a conical cage of micron-sized metallic wires was driven on the MAGPIE pulsed-power facility, which delivers a current ~ 1 MA over ~ 240 ns. The jet formation mechanism relies on the combination of a high rate of radiative cooling together with the redirection of flow across a conical shock. Magnetic fields were not important in these jets and the formation process was purely hydrodynamic. The current experiments are modified in order to introduce a dynamically significant magnetic field in the system.

## NUMERICAL EXPERIMENTS

A radial wire array consists of two concentric electrodes connected radially by 16 tungsten wires 13 µm in diameter and with the radius of the inner and outer electrodes 2 mm and 20 mm respectively. Figure 1 shows the initial set-up of a three-dimensional simulation of a radial wire array. Magnetic field lines are also shown and the regions where the toroidal global field dominates over the local field of the wires can be clearly distinguished. Due to limitations in resolving the micron-sized wire cores in these large scale simulations, the wires are initiated as relatively cold dense gas and not as solid metallic wires. Nevertheless, these artificial initial conditions reproduce correctly the ablation rate of the wires (Lebedev *et al.*, 2001) and the rapid formation of a hot coronal plasma surrounding the wires. The electrodes are treated in the computations as highly conductive but thermally insulated regions. The code solves on an Eulerian grid the three-dimensional single fluid, two temperatures and resistive MHD equations. The evolution of the electromagnetic fields is followed through an explicit Runge-Kutta type time-integration solver and corrected transport. The LTE ionization is calculated using a Thomas-Fermi average atom model, and we also include optically thin radiation as a loss term in the electrons energy equation; the latter is coupled to the energy equation for the ions through an energy equilibration term; more details on code are in Chittenden *et al.*, 2004.

A typical simulation of a radial wire array is shown in Figure 2. The ablation of the wires initially produces an ambient plasma cloud which expands above the plane containing the wires. This thermally dominated plasma provides the collimating environment for the magnetic cavity. When sections of the wire cores are fully ablated the proper magnetic tower jet begins to form, consisting of a magnetic cavity with a jet on its axis. Axial expansion of the cavity and instabilities disrupt the system, leaving a clumpy and collimated jet behind.

The formation of ambient plasma is due to the steady ablation of the wires which produces hot plasma (~ 10 eV) of relatively low resistivity ($\eta$) with respect to the cold (~1 eV) wire cores. For a Spitzer like resistivity $\eta \sim T^{-3/2}$, where T is the temperature of the plasma, a marked difference in the resistivity develops in this two-component structure, with currents preferentially flowing in the ablated plasma. The global magnetic field (see Figure 1) accelerates the ablated plasma in the axial direction, while the wire cores, which are virtually force-free, act as a continuous but stationary source of plasma. We note that resistive diffusion dominates over the advection of the magnetic field up to a height of $l_R$~2-3 mm above the wires, this is approximately the length scale over which the ablated plasma is magnetically accelerated to characteristic velocities of $v_{abl}$~130 km s$^{-1}$. Close to the wires the electron temperature is a few eV and the magnetic Reynolds number $Re_M = v l_R / D_M$ ~ 0.1; $D_M = \eta/\mu_0$ is the magnetic diffusivity and we used $v$~30 km s$^{-1}$. At axial positions above $l_R$ the magnetic Reynolds number increases as a consequence of increasing plasma velocity, temperature and diffusion length scales. Nevertheless, the magnetic field rapidly decays above the wires to ~10% its value calculated in the vacuum region below the wires. The ambient plasma has $\beta$ >1 and the thermal pressure will act to confine the magnetic cavity that forms later. Near the wires the magnetic field pressure dominates and the plasma $\beta$ <1. Over the ablation time ($t_{abl}$~250 ns) a region of height ~30-40 mm above the plane of the wires is filled with plasma, its density varies as ~1/r where $r$ is radial distance from the array's axis. The axially peaked plasma distribution ($n_e$~10$^{18}$ cm$^{-3}$ on axis) occurs as the shock heated radially converging plasma is cooled by radiation losses, resulting in a plasma "column" that is hydrodynamically confined. In the axial direction, because of the time dependent ablation rate (~$I^2$) the density decreases rapidly away from the plane of the wires. Because of the discrete nature of the wires, the plasma distribution is highly modulated in the azimuthal direction (see Figure 3). Nevertheless the evolution of the magnetic cavity is highly symmetric and it is only at later times, as instabilities develop, that asymmetric features become apparent.

Because the mass ablation rate decreases with the strength of the global magnetic field as ~ *1/r* (Lebedev *et al.*, 2001), the highest ablation rate occurs in the proximity of the inner electrode. It is there that during the current discharge full ablation of millimetre-sized sections of the wire cores takes place. Because of the disappearance of the force-free wire cores and thus of the plasma source, the magnetic field pressure associated with the global toroidal magnetic field is now able to sweep the remaining plasma upwards and sideways. The magnetic field acts as a piston, snowploughing the surrounding plasma and forming a magnetic "bubble" inside the background plasma (Figure 3).

In Figure 4 the magnetic field (yellow lines) and the current density (red lines) distribution inside the magnetic cavity are shown for two distinct times. Similarly to the experiments, astrophysical magnetic tower jets are dominated by a toroidal magnetic field which is confined by the pressure of an ambient plasma (Lynden-Bell, 1996, 2003; Kato *et al.*, 2004a; Kato *et al.*, 2004b). With the appearance of the magnetic cavity, a current-carrying jet forms on axis and it is confined by the magnetic field hoop stress. The characteristic density and temperature in the jet are $n_i \sim 3 \times 10^{19}$ cm$^{-3}$ and T $\sim$ 30 eV respectively. The characteristic velocity of the jet is $\sim$ 150 – 200 km s$^{-1}$, which is higher than the initial flow velocity present before the appearance of the magnetic cavity and indicating that the plasma is actually accelerated in the jet formation process. Initially the plasma beta of the jet is $\sim$ 1 and the magnetic Reynolds number is $\sim$ 5-10. With the exception of the jet, the magnetic cavity is mostly void of any plasma. The principal current path is thus along the walls of the magnetic cavity and through the jet (see Figure 4). As noted above, in the jet itself, acceleration of material occurs as plasma swept by the converging magnetic piston, is compressed and redirected axially. In the simulation of tungsten arrays, radiation losses are such that the plasma shell surrounding the magnetic cavity remains fairly thin. Variation of the ambient plasma distribution and the driving magnetic field strength can significantly alter the growth rate of the magnetic tower. The characteristic axial expansion velocity of the magnetic tower is ~ 200 – 400 km/s, while its radial expansion occurs with a velocity of about 50 km/s. The higher velocities are observed for arrays made with the 7.5 µm wires, where the magnetic tower forms during the rise of the current pulse and propagates in an environment having a smaller axial extent. A dominant kink ($m = 1$) mode instability develops immediately after the jet formation and leads to its break-up. For typical jet parameter the growth time (~2.5 ns) of the instability is significantly smaller than the evolution time (Lebedev *et al.*, 2005b). Nevertheless the combination of the axial expansion of the magnetic tower and instabilities do not lead to the destruction of the jet; instead a collimated, clumpy jet is launched out of the cavity. During this transient phase the current and field distribution change significantly (Figure 4): the current begins to connect once again at the base of the magnetic cavity while the magnetic field develops a significant axial component and becomes highly tangled, thus promoting reconnection. Finally, the radiatively cooled, "knotty" jet emerging from the cavity has typical velocities of 200-300 km s$^{-1}$, Mach numbers of >10, plasma beta ~1–10 and Re$_M$ ~1–5. Because of the high Mach number the jet will remain collimated over long distances. In addition, the clumps that form the jet have generally different axial velocity and will interact with each through a series of internal shocks, reminiscent of the internal shocks observed in proto-stellar jets (Hartigan *et al.*, 2001).

The rapid development of instabilities in the jet may be partly suppressed by the presence of a poloidal field in the jet and we are currently developing a series of experiments to investigate its effects. A typical radial array set-up involves the presence of a solenoid-like electrode below the plane of the wires which introduces a longitudinal magnetic field of the order of ~15% of the toroidal field. Field compression, resistive diffusion in the plasma and electrode geometry can all influence the actual topology of the field prior to the jet formation; also the presence of an axial field introduces angular momentum in the flow, further complicating the analysis. Although the exact role of such effects has not yet been clarified, preliminary numerical and experimental results indicate that the inclusion of axial fields and angular momentum can have a major effect on the overall evolution of magnetic towers and on the jet collimation. These results open up the prospect of significantly extending the range of jet studies that can be performed in the laboratory.

**Acknowledgements**

The present work was supported in part by the European Community's Marie Curie Actions – Human Resource and Mobility within the JETSET network under contract MRTN-CT-2004 005592. The authors also wish to acknowledge the SFI/HEA Irish Centre for High-End Computing (ICHEC) and the London e-Science Centre (LESC) for the provision of computational facilities and support.

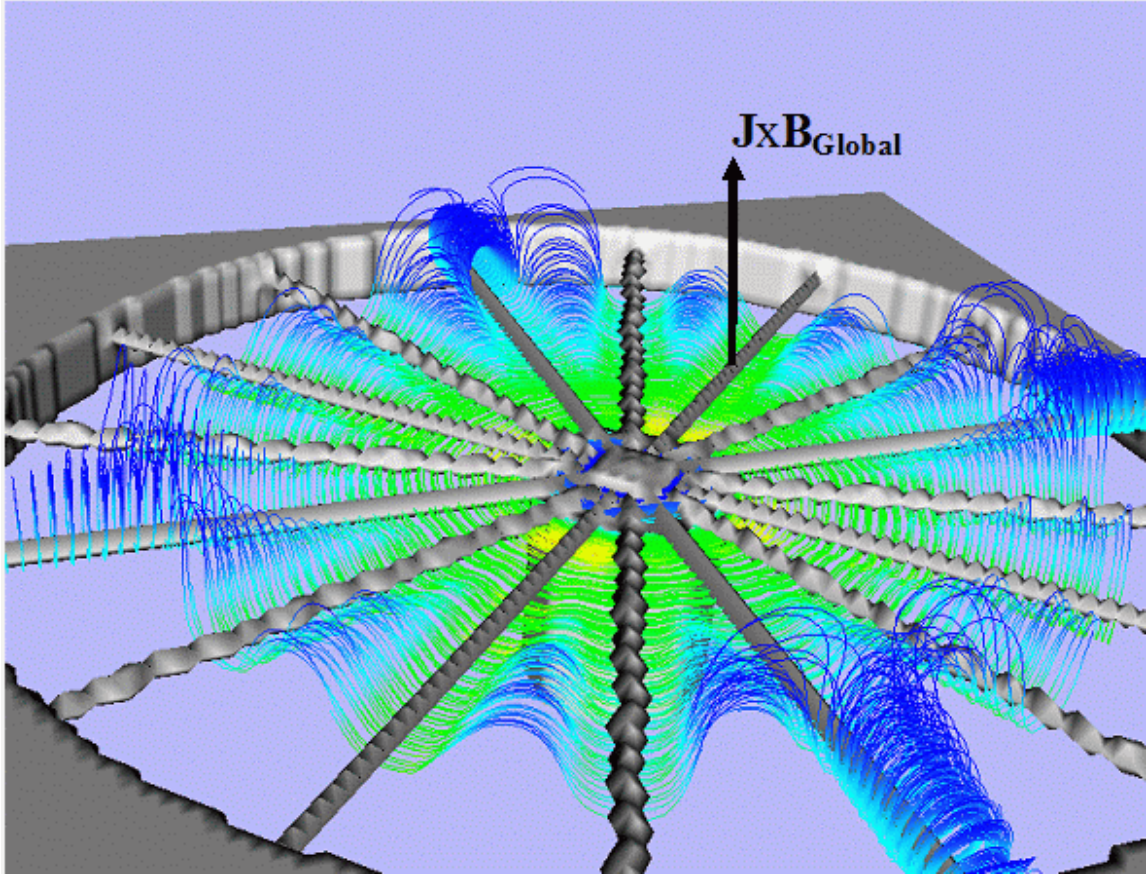

**FIGURE 1**
A radial wire array consists of thin metallic wires connecting two concentric electrodes. The **JxB**$_{Global}$ force accelerates the ablated plasma in the axial direction. The "global" magnetic field, which dominates the system, is purely toroidal. The wires' "private" magnetic field is also plotted for some of the wires.

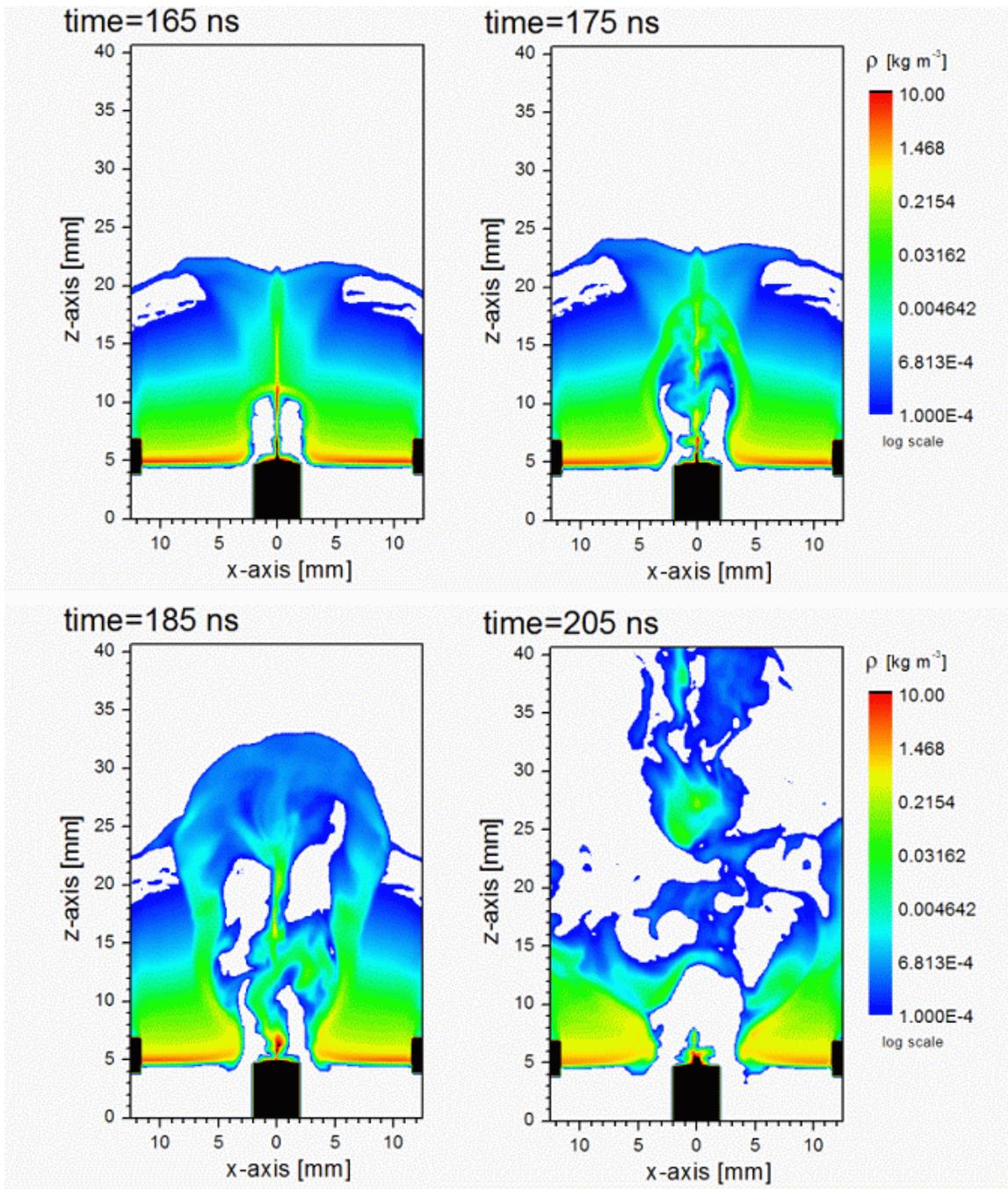

**FIGURE 2**
Time evolution of a radial wire arrays composed of 16 tungsten wires 7.5 µm in diameter. Mass density x-z slices from a 3D simulation are shown. The black square areas indicate the electrodes. Regions where the density is below $10^{-4}$ kg m$^{-3}$ (*white*) are treated as vacuum.

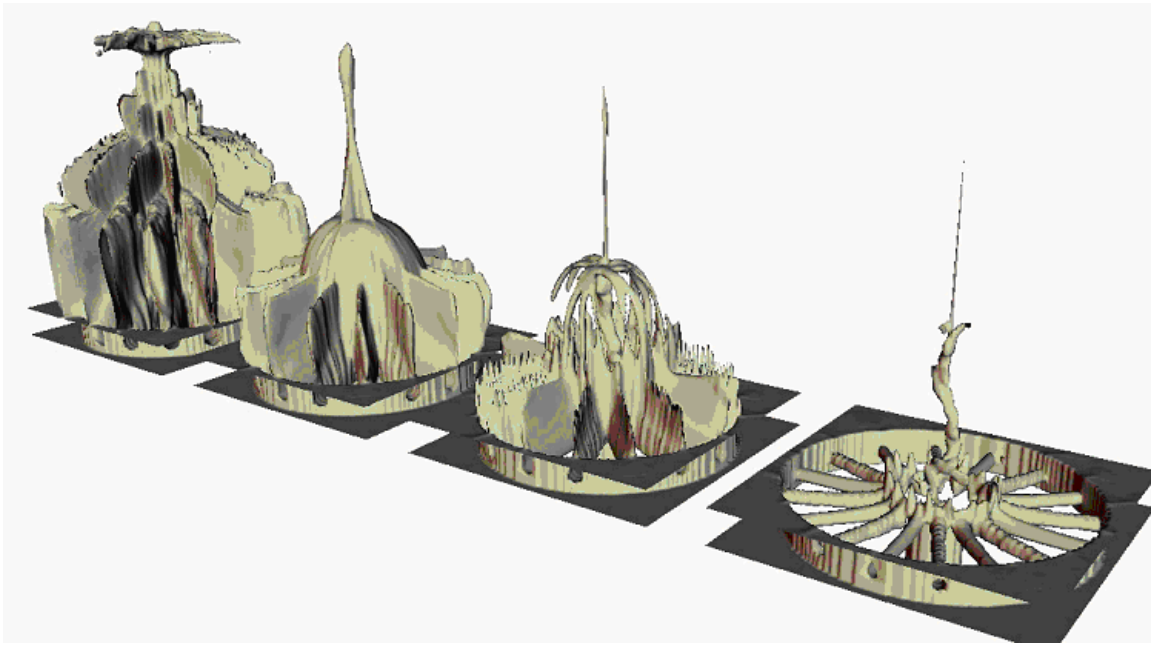

**FIGURE 3**
Four isodensity contours are shown at the same evolutionary time (235 ns). The densities are, from left to right, $5 \times 10^{-4}$, $5 \times 10^{-3}$, $5 \times 10^{-2}$, $5 \times 10^{-1}$ kg m$^{-3}$. The background plasma is visible in the leftmost panel, while in the two mid panels the well developed magnetic cavity can be seen. The rightmost panel shows the jet that forms inside the magnetic cavity.

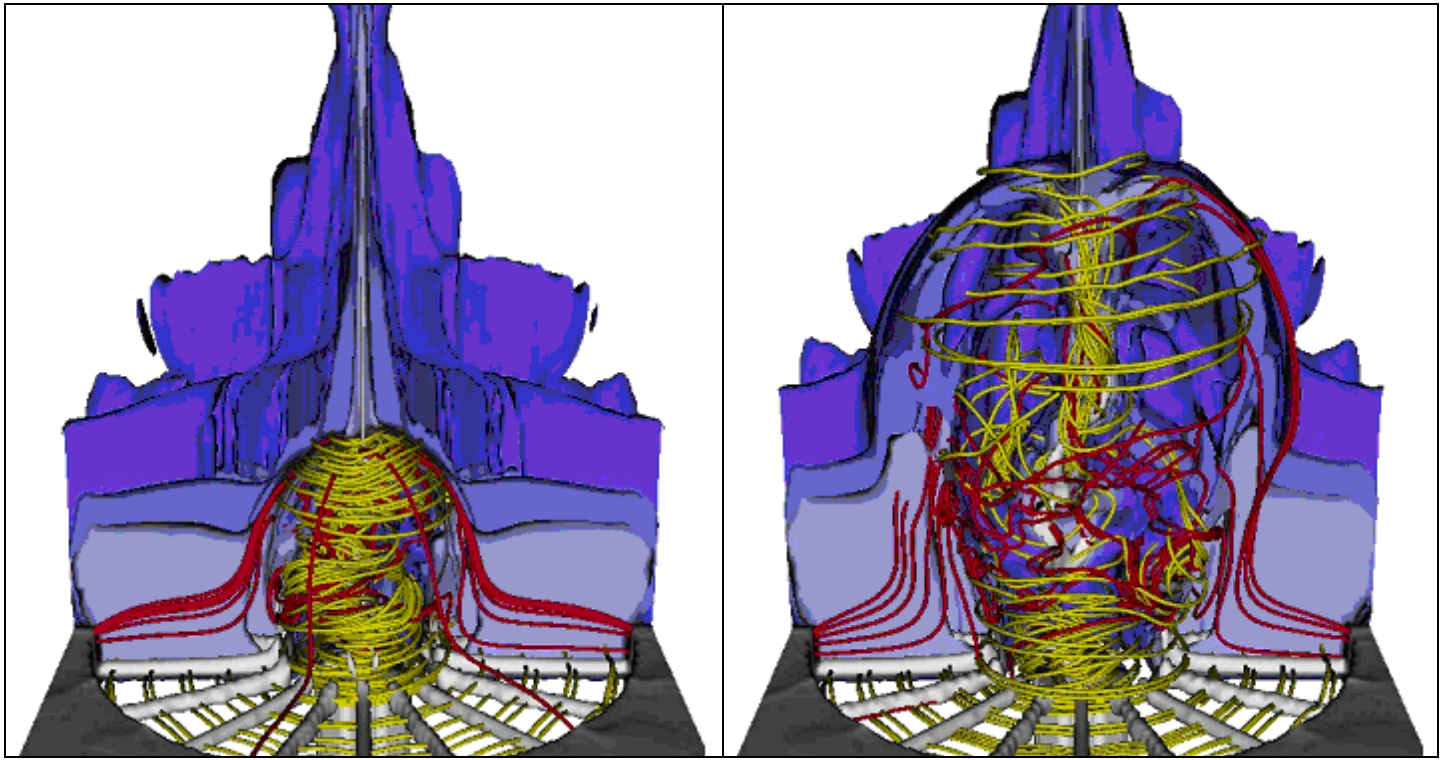

**FIGURE 4** Magnetic field (yellow) and current density (red) distribution inside the magnetic cavity at 225 ns (left) and 245 ns (right). To show the inside of the magnetic cavity the isodensity contours (same as in Figure 3) are sliced vertically.